\documentclass[times,review,nopreprintline]{elsarticle}

\usepackage{framed,multirow}

\usepackage{amssymb}
\usepackage{latexsym}

\usepackage{float}

\usepackage{url}
\usepackage{xcolor}

\usepackage{hyperref}

\usepackage{subcaption}

\usepackage{amsmath}

\usepackage{booktabs}

\usepackage{adjustbox}

\usepackage{tabularx}

\usepackage{pifont}
\newcommand{\cmark}{\ding{51}}%

\definecolor{newcolor}{rgb}{.8,.349,.1}

\journal{Medical Image Analysis}

\begin{document}


\begin{frontmatter}

\title{Overcoming the limitations of patch-based learning to detect cancer in whole slide images}%

\author[1]{Ozan Ciga\corref{cor1}}
\cortext[cor1]{Corresponding author: 
  e-mail: ozan.ciga@mail.utoronto.ca}
\author[5]{Tony Xu\fnref{fn1}}
\fntext[fn1]{This work was conducted while the author was doing internship at Sunnybrook Research Institute.}
\author[3,4]{Sharon Nofech-Mozes}
\author[2]{Shawna Noy}
\author[3,4]{Fang-I Lu}
\author[1,2]{Anne L. Martel}

\address[1]{Department of Medical Biophysics, University of Toronto, Canada}
\address[2]{Sunnybrook Research Institute, Toronto, Canada}
\address[3]{Department of Laboratory Medicine and Molecular Diagnostics, University of Toronto, Canada}
\address[4]{Division of Anatomic Pathology, Sunnybrook Health Science Center, Toronto, Ontario, Canada}
\address[5]{Department of Electrical and Computer Engineering, University of British Columbia, Canada}


\begin{abstract}
Whole slide images (WSIs) pose unique challenges when training deep learning models. They are very large which makes it necessary to break each image down into smaller patches for analysis, image features have to be extracted at multiple scales in order to capture both detail and context, and extreme class imbalances may exist. Significant progress has been made in the analysis of these images, thanks largely due to the availability of public annotated datasets. We postulate, however, that even if a method scores well on a challenge task, this success may not translate to good performance in a more clinically relevant workflow.  Many datasets consist of image patches which may suffer from data curation bias; other datasets are only labelled at the whole slide level and the lack of annotations across an image may mask erroneous local predictions so long as the final decision is correct. In this paper, we outline the differences between patch or slide-level classification versus methods that need to localize or segment cancer accurately across the whole slide, and we experimentally verify that best practices differ in both cases. We apply a binary cancer detection network on post neoadjuvant therapy breast cancer WSIs to find the tumor bed outlining the extent of cancer, a task which requires sensitivity and precision across the whole slide. We extensively study multiple design choices and their effects on the outcome, including architectures and augmentations. Furthermore, we propose a negative data sampling strategy, which drastically reduces the false positive rate (7\% on slide level) and improves each metric pertinent to our problem, with a 15\% reduction in the error of tumor extent.
\end{abstract}

\begin{keyword}
clustering\sep residual cancer burden\sep tumor bed estimation\sep representation learning
\end{keyword}

\end{frontmatter}



\section{Introduction}


Convolutional neural networks (CNNs) are able to extract features from raw images, which has made them attractive in many visual tasks, including medical image analysis. In computational pathology, current approaches mostly focus on isolated patches extracted from WSIs, or tasks that are tailored around a slide or patient-level predictions, as opposed to requiring consistently high precision and sensitivity across a WSI \citep{srinidhi2019deep}. We argue that applying CNNs on WSIs might not achieve similar performances due to several biases outlined below, in addition to the different characteristics of WSIs and patches extracted from WSIs. 

\begin{figure}[H]
     \centering
     \begin{subfigure}[b]{0.45\textwidth}
         \centering
         \includegraphics[width=1\textwidth]{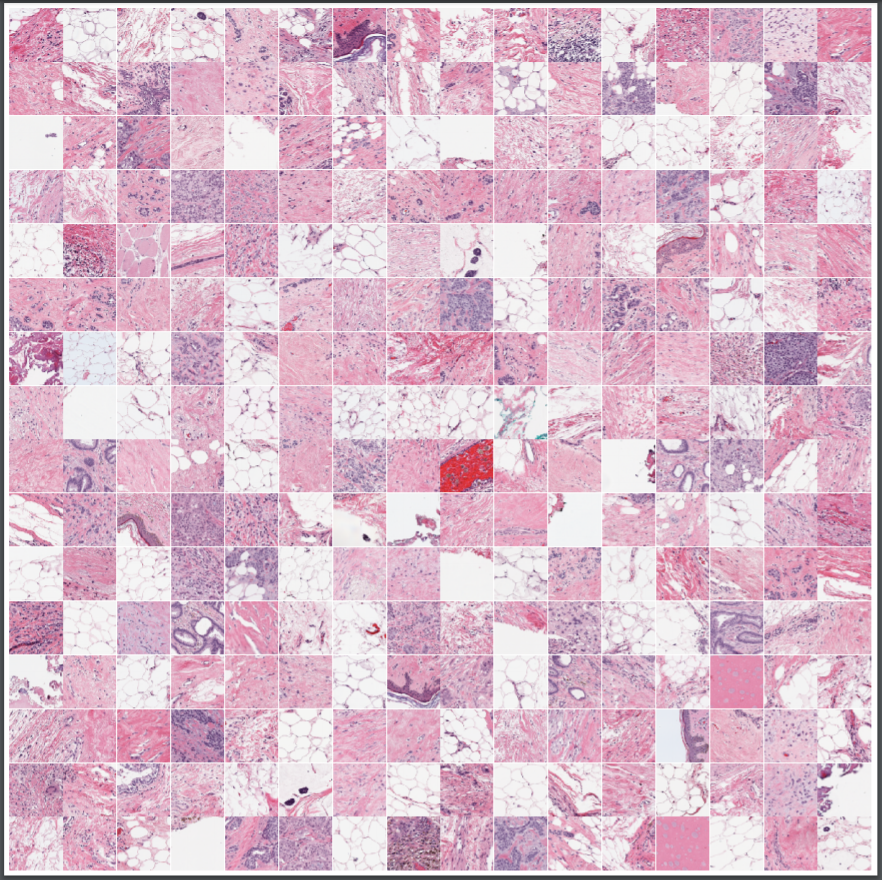}
         \caption{Random sampling}
         \label{fig:neg_sampling_random}
     \end{subfigure}
     \begin{subfigure}[b]{0.5\textwidth}
         \centering
         \includegraphics[width=1\textwidth]{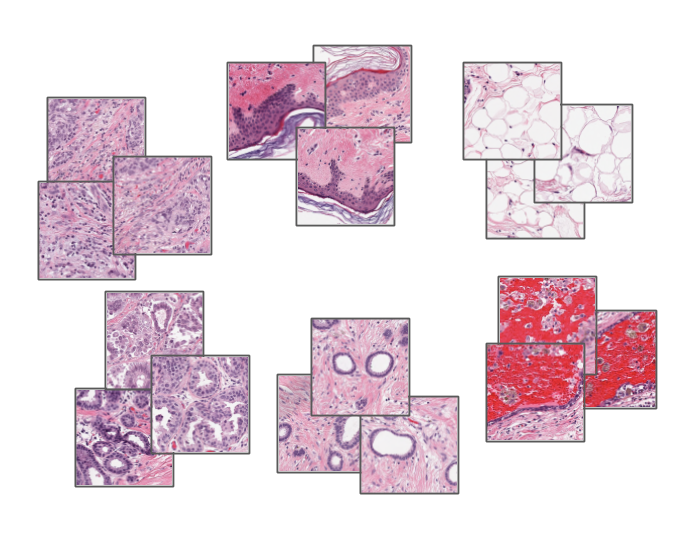}
         \caption{K-means sampling}
         \label{fig:neg_sampling_kmeans}
     \end{subfigure}
        \caption{Negative mining with random sampling versus feature based K-means clustering from whole-slide images. Both figures are extracted from actual experiments. In Fig. \ref{fig:neg_sampling_kmeans}, we show the three samples closest to the centres of six randomly selected clusters for illustration. }\label{fig:neg_sampling}
\end{figure}

Much of the early work in digital pathology involved the extraction of a small fraction of patches from each WSI, which were then annotated or labelled by pathologists.  This approach is vulnerable to a data curation bias which refers to the implicit assumptions made by curators that may affect the task outcome. In patch-level problems, training and validation datasets are generally collected by the same experts or with the same instructions. For example, both training and validation sets contain the same number of unique classes with similar ratios. Data curation is also generally biased towards the positive class since these images are usually collected from patients undergoing treatment. Curation bias is exploited for tuning the network hyperparameters (e.g., the confidence threshold) on validation sets to maximize performance on the held-out test set. While these are appropriate in the context of a challenge, they do not necessarily hold for tasks with WSIs, where it is not possible to determine a universally best threshold or to comprehensively annotate data (e.g., representing all negatives in cancer detection, such as ink, creases, out-of-focus regions or non-cancerous tissue). 

 Although CNNs can only be applied to individual patches within a WSI, slide level tasks can be performed by aggregating the individual patch level results.  It is, however, important to take into account how this aggregation is carried out when evaluating and comparing algorithms. Many tasks that involve WSIs are binary decision problems or allow a number of errors without negatively impacting the measured metric. For instance, \cite{Bandi2018} aim to determine the severity of breast cancer by measuring the extent of metastasis via a five-class problem. However, if a single macro-metastasis is identified, the exact number of metastases become irrelevant for certain classes by definition. Similarly, \cite{campanella2019clinical} aim to determine whether cancer is present or not present in a WSI. On slides labelled as containing cancer, false positive are not penalized because the location of the detected cancer is not important; conversely, in slides labelled as negative, there is no differentiation between one or many false positive detections. In contrast, the number of errors at the patch level will be related  directly to task performance in slide level tasks that involve quantification or segmentation.  
 
CNNs usually work best with small images such as $224\times224$ pixels, whereas a WSI can be $100,000\times100,000$ pixels. Even a drastic down-sampling (e.g., 16$\times$) will necessitate tiling and working on a portion of the WSI. Modifying the architecture, or simply feeding a larger image to overcome this issue is not always possible due to optimization challenges and graphical memory constraints. While the latter is less prohibitive and may be mitigated by technological advancements, training models with more parameters to accommodate WSIs with larger convolutional kernels are infeasible due to $\mathcal{O}((\textrm{kernel side length})^2)$ scaling. This upper bounds the performance on WSIs since CNNs heavily rely on context and edges, which may not always be present, especially when the WSI is viewed under high resolution (e.g., the outline of a gland or duct may only be partially available if the images are sampled at high resolution over a limited field of view).

In this paper, we extensively study the differences between WSI analysis and patch level analysis. We use WSIs of breast tissue which has been surgically resected following neoadjuvant therapy (NAT) to explore the differences between the following three tasks: 
\begin{enumerate}
\item Patch-level classification: determine whether cancer cells are present or not in a curated set of patches from WSIs \citep{Martel2019tcia}.  
\item Slide-level classification: determine whether a WSI contains cancer cells or not without localizing the cancer within the WSI.
\item Slide-level segmentation: Find the convex hull that contains all cancer cells in a WSI  (i.e., the tumor bed estimation).
\end{enumerate}

Importantly, a binary classification network is used in each task. For the segmentation task, the WSI is tiled into patches, and each patch is classified as cancer or no cancer prior to computing the convex hull. We observe that the best practices, including the most suitable augmentations and the appropriate network complexity, are different for each task. We find a network achieving $>99\%$ F1 score on patch-level classification fails to identify the tumor bed accurately, with a Dice score of $\sim 50\%$. We then study failure points of the patch-level classifier and propose a novel negative sampling method that improves the task performance beyond random or no sampling. We also observe a large gap between slide-level classification and segmentation. For the former, we achieve $94\%$ accuracy, whereas for the latter, we obtain a Dice score of $75\%$.





\section{Problem definition}


To exemplify the differences between patches and WSIs, we tackle the automatic detection of breast cancer. Breast cancer is the second most common invasive cancer in women, where around 12\% of women having a lifetime risk of getting breast cancer \citep{desantis2014breast}. New therapies have improved outcomes and it is increasingly important to assess the stage and subtype of cancer more precisely in order to select the most appropriate treatment. While tools that can  identify and characterize cancer in a specific region of the tissue specimen selected by a pathologist have value, there is a need for automated tools capable of  processing  WSIs since cancer may be anywhere within a slide, or even spread across multiple slides.

An evaluation of the residual tumour in surgically resected breast tissue after neoadjuvant therapy for breast cancer is clinically relevant in determining treatment response and estimating prognosis. It has been shown that the extent of the residual tumor bed, the cellularity within the tumor bed, and the presence or absence of lymph node matastases can be used to estimate the five-year recurrence risk  \citep{symmans2007measurement, symmans2017long}. While automated analysis of WSIs for breast cancer tasks is common \citep{Bejnordi2017, Bandi2018, aresta2019bach}, there have been few studies that focus on post-NAT images or slide-wide analysis \citep{akbar2019automated}. Recently, \citep{campanella2019clinical} conducted a large scale study with 9894 slides from axillary lymph nodes for breast cancer metastasis detection, and achieved 0.989 area under the curve (AUC). 52\% of the false negatives were from slides that showed signs of neoadjuvant chemotherapy, indicating the challenges associated with identifying cancer in post neoadjuvant therapy images, even when the dataset size is large. In addition, 8.6\% of false negatives were due to isolated single tumor cells, which play an important role in determining the tumor bed outline as they may change the extent significantly.

Due to its non-trivial nature, clinical relevance and sensitivity towards both false positives and negatives, we conduct experiments for the tumor bed estimation  task on WSIs of resected breast tissue. The tumor bed is defined as the convex hull that contains all residual cancer cells and the task requires that all residual cancer cells are detected; after therapy this is more difficult as isolated cancer cells may be surrounded by stromal tissue. The task also requires a very low false positive detection rate as the erroneous detection of cancer cells can lead to large errors in the boundaries of the tumor bed.



\section{Methods}

\subsection{Data}

The BreastPathQ training dataset \citep{akbar2019automated} consists of 2394 patches (of size 512$\times$512) extracted from 69 WSIs of Post-NAT-BRCA specimens \citep{peikari2017}. BreastPathQ images are given labels according to the percentage of cellularity in each patch. The training set is imbalanced, where 71\% of the images are tumor-positive. The corresponding WSIs \citep{Martel2019tcia} also include tumor bed outlines for each slide which we use for sampling additional negative patches from tumor-free regions of the WSIs.

We use the testing images from the BreastPathQ challenge for the patch-level classification task. This set includes 1106 images, and is annotated by two independent pathologists. The test set for the slide-level classification and segmentation tasks is composed of a private set of 50 WSIs, evenly and randomly sampled from 10 independent cancer patients in different treatment stages. Three pathologists have independently annotated tumor bed outlines in each WSI and later reviewed each WSI together to create a consensus ground truth; this allows us to assess the inter-expert disagreement rate. In our experiments, we use the consensus ground truth unless otherwise stated. According to the consensus dataset, 30\% of the WSIs are tumor-free.



\subsection{Preprocessing}

Regions without nuclei are not relevant in identifying tumors. Therefore, we threshold each WSI in HSV color space to remove such regions, with $0.65>hue>0.5$, $saturation>0.1$, and $0.9>value>0.5$. These bounds are experimentally determined using a separate public dataset for breast cancer segmentation \citep{aresta2019bach}. We tile the WSI and pass patches with foreground ratios $\geq$ 25\% through the model during the evaluation stage.

\subsection{Patch-level classification task}

We use the BreastPathQ dataset for patch-level classification. The continuous cellularity percentage values per patch are mapped to two classes, with cellularity percentages above zero labelled as cancer.

\subsection{Slide-level tasks}

In addition to the same dataset used for patch-level classification, we incorporate additional negative (no cancer) samples for the slide level tasks (see Section \ref{sec:neg_sampling} for details). 

Our method achieves slide-level segmentation (tumor bed estimation) by tiling the WSI into distinct patches and generating a probability value (of being tumor-positive) per patch. One may slide a window in strides smaller than the patch length to improve the classification of each pixel through a voting scheme. For example, given a patch size of 512, a stride of 256 will generate four separate probability values per pixel, representing if the patch contains any tumor. Then, these values may be averaged to obtain more accurate probability values per pixel. Since each value is obtained using a different patch, striding helps to incorporate context better.

The predicted label for the slide-level classification can be obtained from the segmentation output. Specifically, if the tumor bed exists, or there is one or more positive patch on the WSI, the WSI is labeled as tumor-positive. 

\subsection{Negative sampling}\label{sec:neg_sampling}

A single WSI may contain artifacts (e.g., creases, random blurring of regions, ink or staining residues), as well as structures that are considered as background (e.g., histocytes, adenosis, or red blood cells in tumor bed estimation) which may not have been filtered out in the preprocessing step. While all datasets provide negative samples, it is not feasible to label all forms of negatives. Therefore in digital histopathology, it is common for the negative (e.g., healthy or benign tissue in breast cancer detection) class to be underrepresented. 

BreastPathQ dataset consists of 69 WSIs, where each WSI is annotated for the tumor bed outline. By definition, patches extracted from outside of the tumor bed region are considered as completely cancer-free. We sample patches from outer extents of each WSI and collect a total of 1.4 million images. Then, we sample 21 thousand images randomly or by clustering. For clustering, we use features ($\in \mathbb{R}^{1280}$) generated by EfficientNet-B0 model trained on ImageNet  \citep{efficientnet}. Specifically, we use features corresponding to each image obtained by 2d-average pooling of the final layer of the network prior to the fully-connected classification layer (also called as the pre-activation). We then cluster 1.4 million feature vectors into three thousand clusters using the mini-batch K-means algorithm and select seven instances from each cluster. Examples of sampled images from both random and K-means sampling strategies can be viewed in Fig. \ref{fig:neg_sampling}.


\section{Experiments}

\subsection{Experimental setup}

For all experiments, we use the EfficientNet framework \citep{efficientnet}, and train the network for predicting if a patch of size $224\times224$ (training patches are of size $512\times512$, which are randomly cropped to $448\times448$ before resizing each dimension in half) contains cancer cells. EfficientNet is based on AutoML framework and compound scaling, and achieves better performance than standard architectures such as residual networks with a fewer number of trainable parameters. We use Adam optimizer with $\beta_1 = 0.9$, $\beta_2 = 0.999$, learning rate of 0.0001, batch size of 20, and weighted cross-entropy loss function, and train for 250 epochs for each experiment. We split 15\% of the training data for validation, and select the model with minimum validation loss in testing. Unless otherwise stated, we use the EfficientNet-B0 model. 

\subsection{Metrics}

\begin{figure*}
    \centering
     \begin{subfigure}[t]{0.245\textwidth}
         \includegraphics[width=\textwidth]{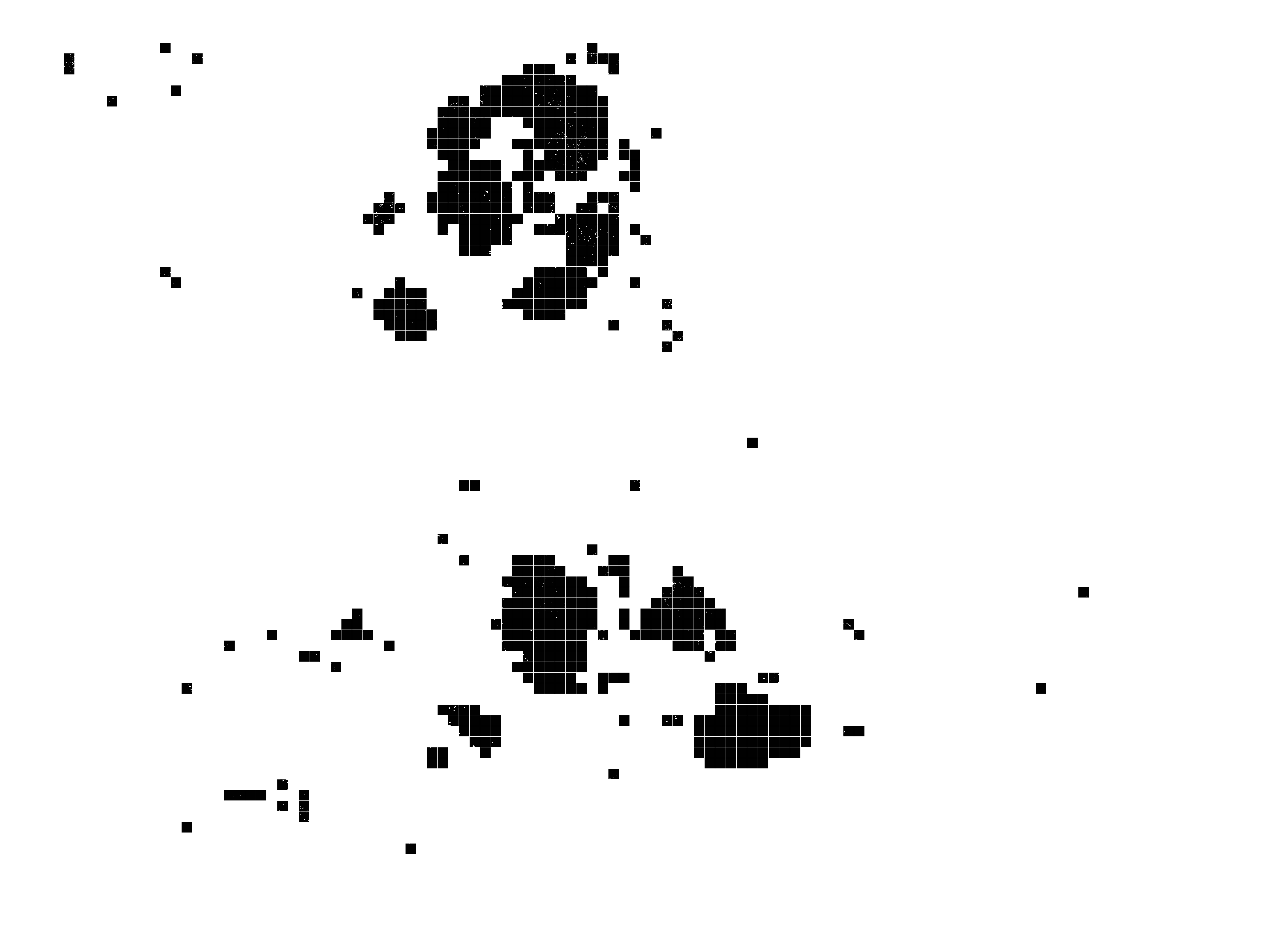}
         \caption{The heatmap prediction output from the network}\label{fig:postprocess_a}
     \end{subfigure}
     \begin{subfigure}[t]{0.245\textwidth}
         \includegraphics[width=\textwidth]{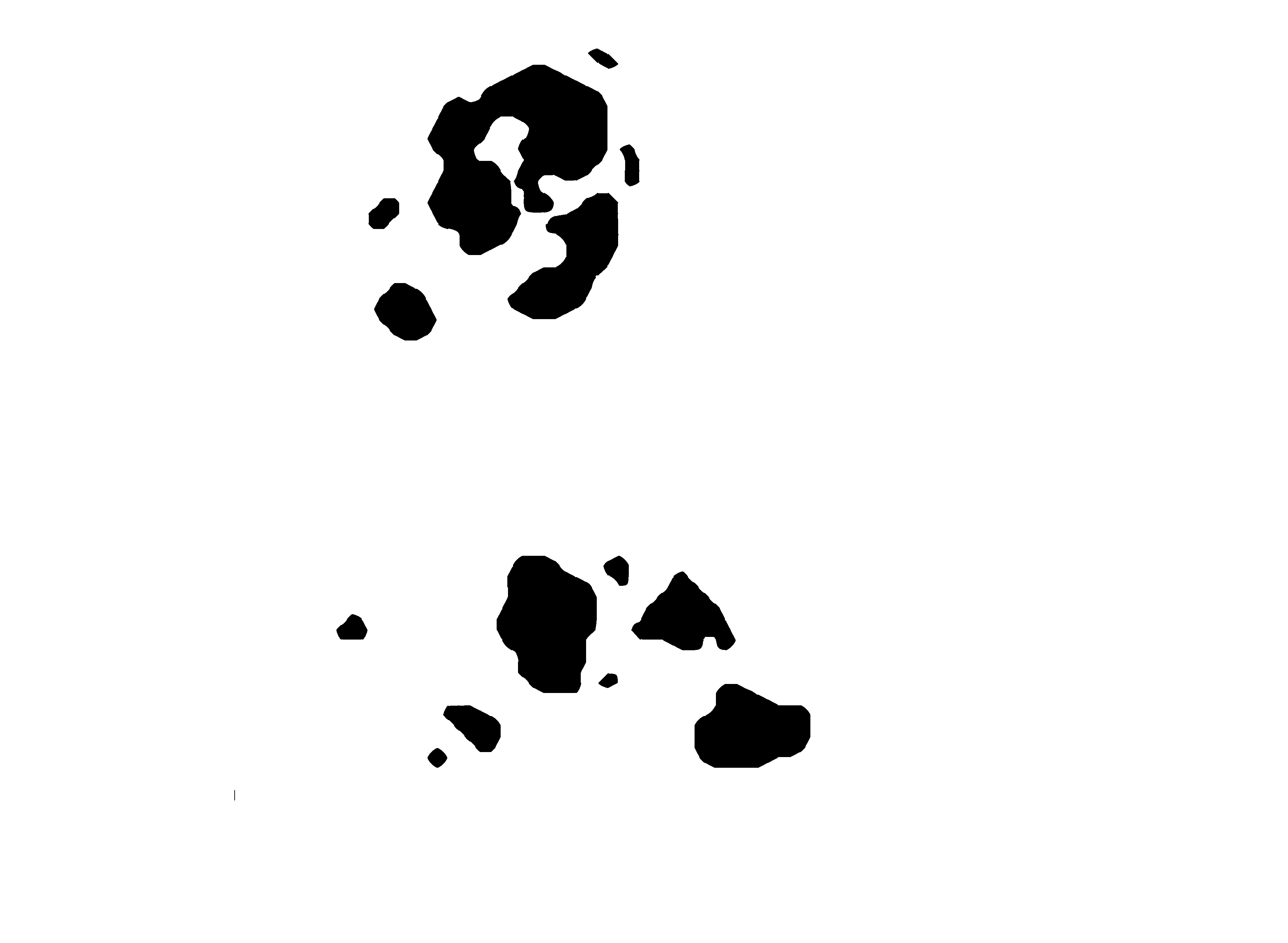}
         \caption{Binary thresholding of the prediction heatmap}\label{fig:postprocess_b}
     \end{subfigure}
     \begin{subfigure}[t]{0.245\textwidth}
         \includegraphics[width=\textwidth]{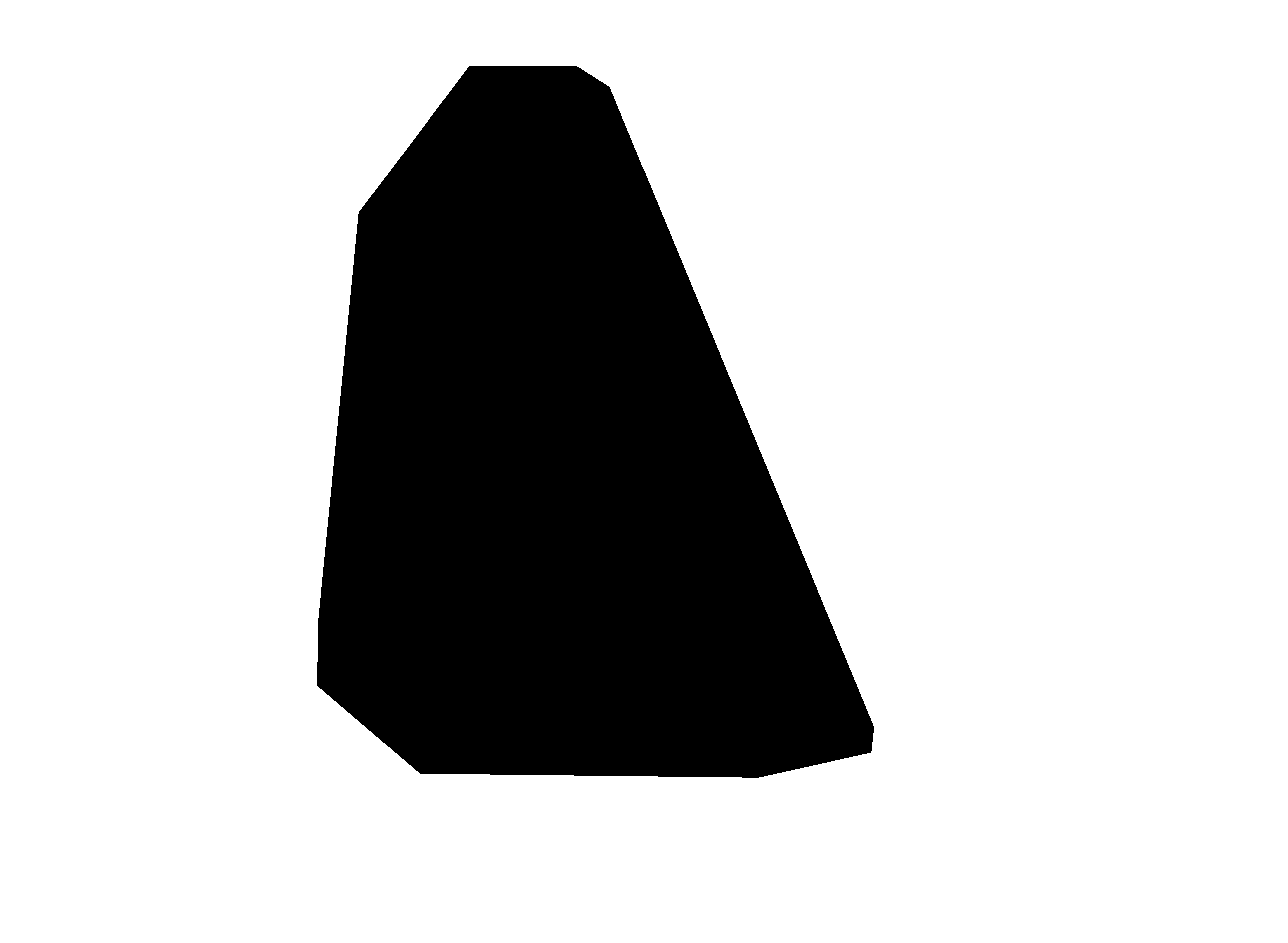}
      \caption{The convex hull of the predicted binary mask}\label{fig:postprocess_c}
     \end{subfigure}
     \begin{subfigure}[t]{0.245\textwidth}
         \includegraphics[width=\textwidth]{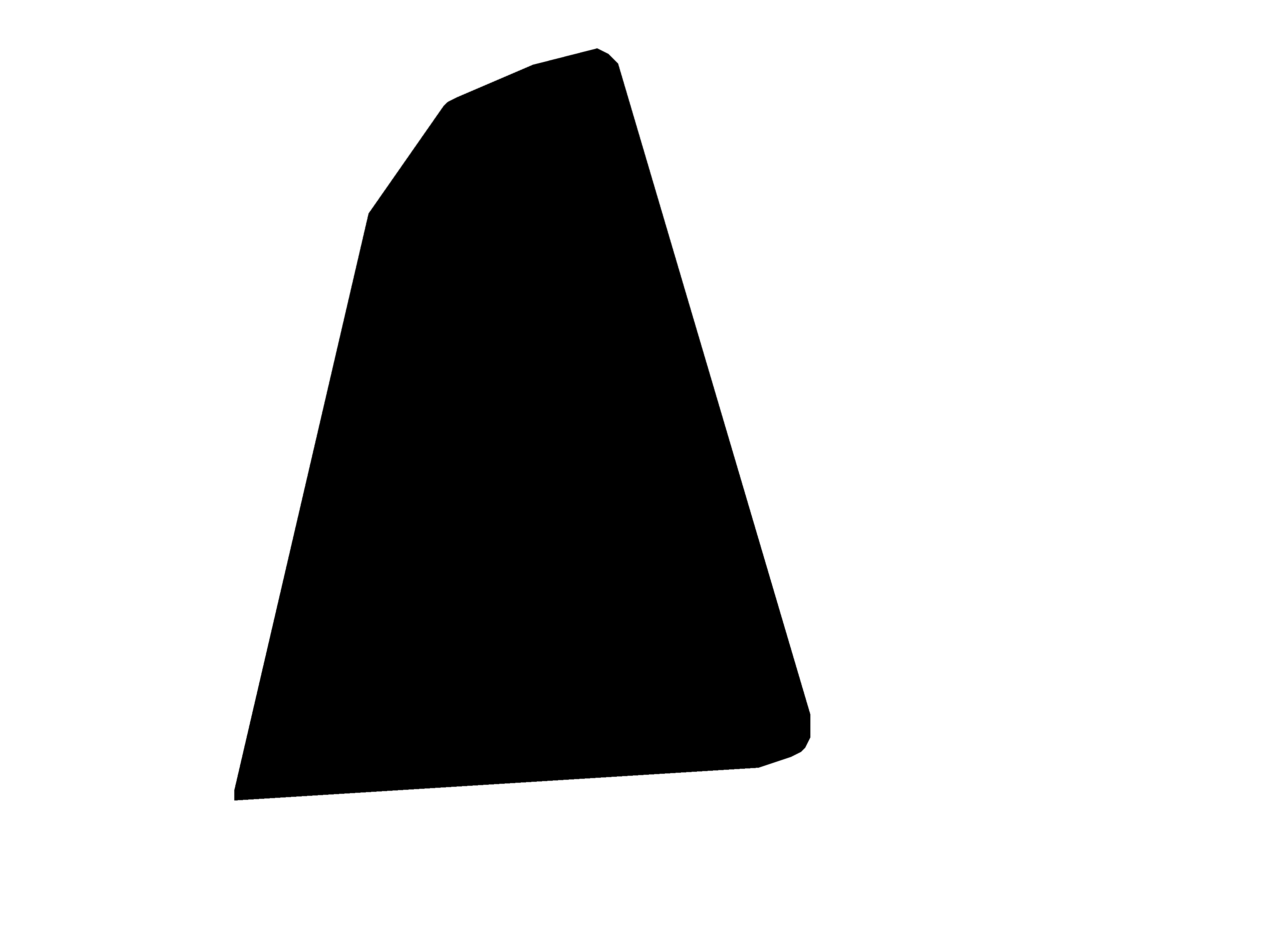}
     \caption{Ground truth for the image}\label{fig:postprocess_d}
     \end{subfigure}
     \par
     \begin{subfigure}[t]{0.245\textwidth}
         \includegraphics[width=\textwidth]{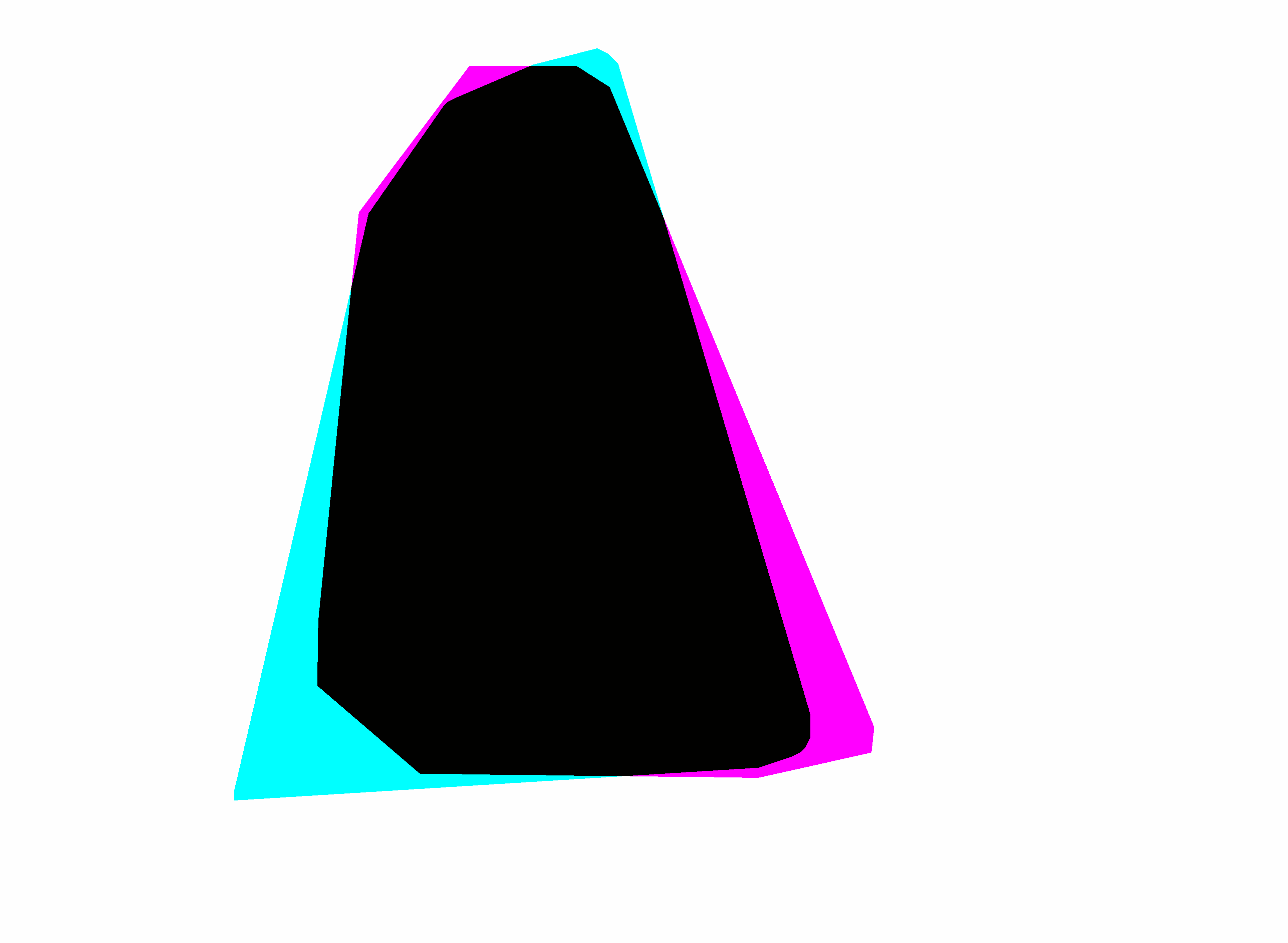}
         \caption{Overlap between ground truth and the prediction (black), and the discrepancies (magenta and cyan)}\label{fig:postprocess_f}
     \end{subfigure}
      \begin{subfigure}[t]{0.245\textwidth}
     \includegraphics[width=\textwidth]{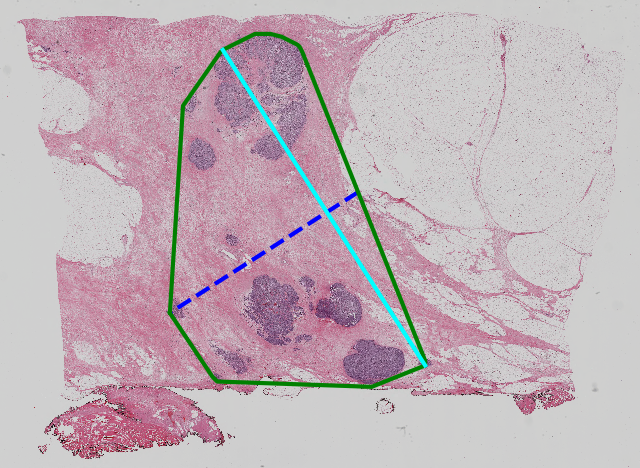}
     \caption{Tumor bed extent estimation of the predicted mask.}\label{fig:postprocess_h}
 \end{subfigure}
      \begin{subfigure}[t]{0.245\textwidth}
         \includegraphics[width=\textwidth]{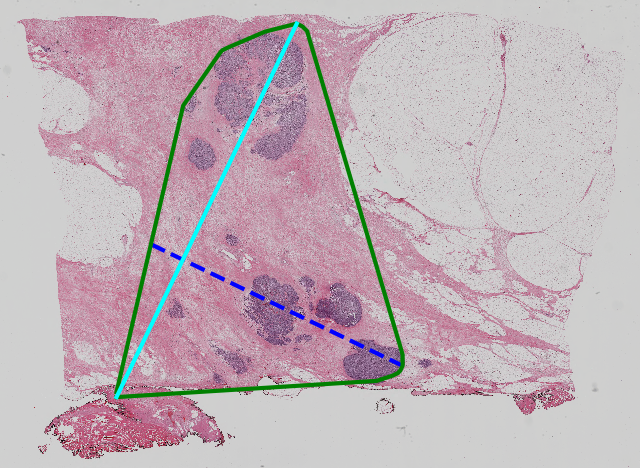}
         \caption{Tumor bed extent estimation of the ground truth mask}\label{fig:postprocess_g}
     \end{subfigure}
       \begin{subfigure}[t]{0.245\textwidth}
     \includegraphics[width=\textwidth]{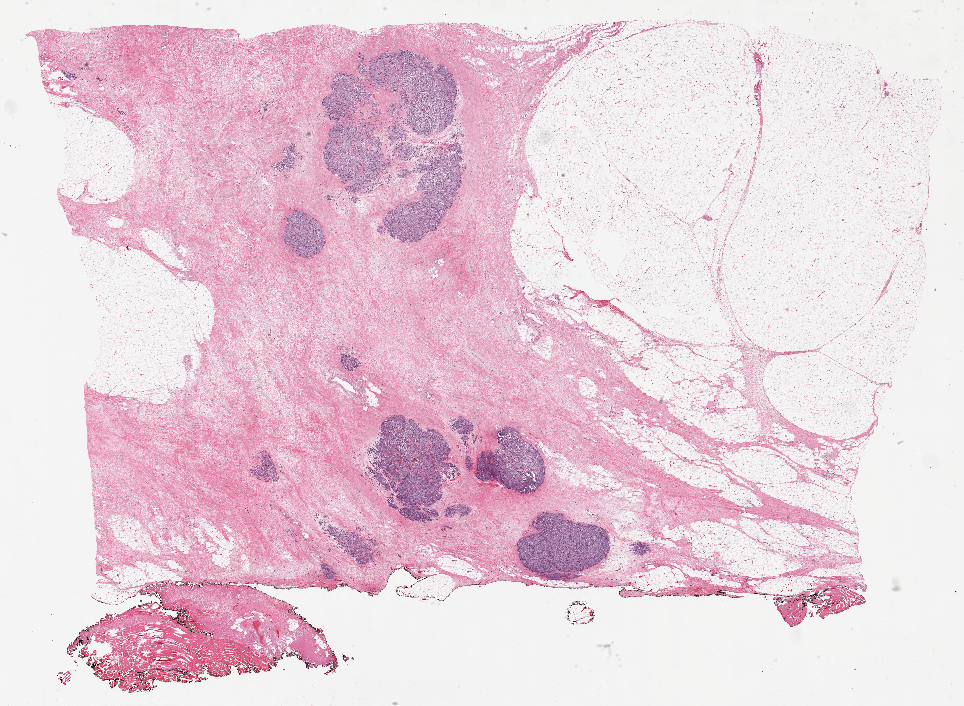}
     \caption{Image used for the demonstration for figures \ref{fig:postprocess_a} - \ref{fig:postprocess_h}}\label{fig:postprocess_i}
     \end{subfigure}
        \caption{Comparing overlap (Dice coefficient) and difference between ground truth and the prediction in $d_{prim}=\sqrt{d_1 d_2}$, where $d_1$ is the primary or the longest (turquoise), and $d_2$ is the secondary (blue) diagonal. The process is visualized from left to right, then top to bottom (row major order).}
        \label{fig:hsv_threshold}
\end{figure*}

In binary decision problems, such as determining \textit{any} presence of tumor in WSIs, finding a single positive region suffices for identifying the positive WSIs, and any additional false positives or false negatives for the same WSI do not incur any penalty. In contrast, our task requires us to be consistently precise throughout the WSI in order to estimate the tumor bed accurately. Therefore, we report both the confusion matrix to quantify if our system is able to detect tumor accurately or discard slides without tumor, as well as the error in $d_{prim}$ (see Fig. \ref{fig:hsv_threshold} for visualization and \ref{apx:tumor_bed_calc} for details), defined as the absolute difference between the prediction and ground truth averaged over the test set, and the overlap (measured by the Dice coefficient) between the ground truth and the prediction. Note that $d_{prim}$ is especially important in assessing the quality of predictions since, unlike the Dice coefficient, this metric is very sensitive to distant false positives and false negatives towards the WSI's outer edges. 

For our purposes, the confusion matrix is related to the slide-level classification task, whereas the Dice coefficient and $d_{prim}$ are related to the slide-level segmentation task. Importantly, segmentation-related metrics are not calculated for WSIs where no tumour is detected. This means that a lower error in $d_{prim}$ may be due to the exclusion of WSIs that are falsely classified as negative for residual tumour when calculating the average value.  

\subsection{Tasks}

\subsubsection{Patch-level cancer detection}

\begin{table}
\caption{Binary (tumor vs. no tumor) classification accuracies on BreastPathQ test set with different negative sampling schemes using minority oversampling.}
\centering
\begin{tabularx}{0.6\linewidth}{X >{\hsize=.85\hsize}X X}
\toprule
Negative sampling & Acc. & Cfs. mtx. \\
\midrule
K-means & 92.6\% & $\begin{pmatrix} 173 & 58 \\ 23 & 852 \end{pmatrix}$ \\ 
Random & 96.3\% & $\begin{pmatrix} 200 & 31 \\ 9 & 866 \end{pmatrix}$ \\ 
None & 98.7\% & $\begin{pmatrix} 224 & 7 \\ 7 & 868 \end{pmatrix}$ \\ 
\bottomrule
\label{tab:patch_level_cls}\end{tabularx}
\end{table}


We use models trained with minority oversampling weighing (see Table \ref{tab:negative_sampling}), and assess their performance on patch-level BreastPathQ challenge test set with 1106 samples. We also conduct experiments to assess the effects of augmentation, size of the architecture and number of test time augmentations. 


\subsubsection{Whole-slide image cancer detection}

After separately annotating the test set, pathologists reviewed the dataset together and discarded four slides due to high disagreement rates. In these slides, pathologists disagreed as to the presence of tumor, annotated non-overlapping regions as the tumor bed, or disagreed on highly contentious regions where the presence of tumor could not be determined by only using H\&E stained slides. For the remaining slides, a consensus tumor bed was selected using a combination of expert annotations. Any additional regions were not examined during this step.

\subsection{Results}

\subsubsection{Patch image analysis}


\begin{table}[h]
\caption{Binary (tumor vs. no tumor) classification accuracies on BreastPathQ test set under different settings. Unless otherwise stated, test-time augmentations (TTA) are set to two. }
\centering
\begin{tabularx}{\linewidth}{X X X}
\toprule
Setting & Acc. & Cfs. mtx. \\
\midrule
No augmentation & 93.5\% & $\begin{pmatrix} 186 & 45 \\ 27 & 848 \end{pmatrix}$ \\ 
Stain norm & 92.4\% & $\begin{pmatrix} 180 & 51 \\ 32 & 843 \end{pmatrix}$ \\ 
Color jittering & 93.5\% & $\begin{pmatrix} 205 & 26 \\ 46 & 829 \end{pmatrix}$ \\ 
Affine trs. & 94.4\% & $\begin{pmatrix} 207 & 24 \\ 38 & 837 \end{pmatrix}$ \\ 
\midrule
TTA = 1 & 92.9\% & $\begin{pmatrix} 198 & 33 \\ 45 & 830 \end{pmatrix}$ \\ 
TTA = 2 & 93.5\% & $\begin{pmatrix} 186 & 45 \\ 27 & 848 \end{pmatrix}$ \\ 
TTA = 3 & 93.3\% & $\begin{pmatrix} 183 & 48 \\ 26 & 849 \end{pmatrix}$ \\ 
TTA = 4 & 92.9\% & $\begin{pmatrix} 178 & 53 \\ 25 & 850 \end{pmatrix}$ \\ 
TTA = 16 & 92.2\% & $\begin{pmatrix} 167 & 64 \\ 22 & 853 \end{pmatrix}$ \\ 
\midrule
B0 & 93.5\% & $\begin{pmatrix} 186 & 45 \\ 27 & 848 \end{pmatrix}$ \\ 
B3 & 93.1\% & $\begin{pmatrix} 189 & 42 \\ 34 & 841 \end{pmatrix}$ \\ 
B5 & 95.1\% & $\begin{pmatrix} 201 & 30 \\ 24 & 851 \end{pmatrix}$ \\ 
B7 & 93.5\% & $\begin{pmatrix} 193 & 38 \\ 34 & 841 \end{pmatrix}$ \\ 
\bottomrule
\label{tab:patch_level_cls_all}\end{tabularx}
\end{table}

We compare three different negative sampling methods in Table \ref{tab:patch_level_cls}. The negative instances are sampled from the WSIs as described in Section \ref{sec:neg_sampling}. Minority oversampling performed the best for all sampling strategies. We compare multiple augmentation strategies, number of test time augmentations (TTA), and the effect of model complexity on the classification performance in Table \ref{tab:patch_level_cls_all}.

\subsubsection{Whole slide image analysis}\label{sec:results_slide_level}

The results of our experiments are presented in tables \ref{tab:negative_sampling}, \ref{tab:data_augmentation}, \ref{tab:model_complexity} and \ref{tab:stride_length}. Note that the average ground truth $d_{prim}$ is 14.5 millimeters, with a standard deviation of $6.1$. We compare the discrepancy between references (i.e., the pathologists) to each other, as well as to the network in Table \ref{tab:ref_target_consensus} after the consensus set (with 46 slides) has been established. We also make the same comparison prior to consensus, accounting for all 50 slides in Table \ref{tab:ref_target_beforeconsensus}.

We first determine the best negative sampling method, along with options for weighing different samples depending on their class (Table \ref{tab:negative_sampling}). We compare three different sampling strategies, including the clustering approach defined in Section \ref{sec:neg_sampling}, randomly selecting negative instances as well as no additional negative sampling beyond what the training data already has. Negative mining creates an imbalance in favor of the negatives (85\% of the training data). Therefore, we use multiple balancing strategies, including multiplying the incurred loss with a coefficient that is directly ($\propto$) or inversely ($\frac{1}{\propto}$) proportional to its corresponding class. We also conduct experiments using no balancing ($=$) and balancing by oversampling. For our training data, each negative sample is passed through the network two times, whereas each positive sample is passed 11 times per epoch during the training.

We also compare the effect of different types of image augmentation on slide-level classification and segmentation tasks (Table \ref{tab:data_augmentation}). We compare randomly cropping $448\times448$ rectangular boxes from the training instances versus no cropping both during training and evaluation. We modify the color properties of each instance by applying color jittering, which amounts to randomly changing brightness, contrast, saturation and hue values by 5\% from their original values, as well as by stain normalization using a reference staining matrix \citep{macenko2009method}. We experiment applying affine image transformations by scaling the cropped patch down or up by 10\%, and shearing by 60$^{\circ}$ in both x and y axes.

We compare four different models with varying widths and depths to understand the effect of model complexity on slide-level task outcomes (Table \ref{tab:model_complexity}). We also compare multiple sliding window approaches for finding the most suitable tile stride (Table \ref{tab:stride_length}). Finally, we compare our method to each of the three pathologists, along with comparing pathologists to each other in tables \ref{tab:ref_target_consensus} and \ref{tab:ref_target_beforeconsensus}.

\begin{table}
\caption{The effect of sampling negative instances randomly versus a clustering approach. Under each sampling strategy, we compare various methods to handle class imbalance. $\propto$ indicates using class weights in cross-entropy loss directly proportional to class distributions, $=$ uses weights 1 for all classes. Each setting uses the EfficientNet-B0 framework, 256 as the tile stride, and only random cropping as data augmentation.}
\begin{tabularx}{\linewidth}{X | X X X X}
\toprule
Negative sampling & Weighing & Dice coefficient & Error in $\sqrt{d_1 d_2}$ & Confusion matrix \\
\midrule
K-means & $\propto$ & 0.68 & 2.82 & $\begin{pmatrix} 7 & 7\\ 2 & 30 \end{pmatrix}$ \\ 
 & $\frac{1}{\propto}$ & 0.70 & 3.10 & $\begin{pmatrix} 4 & 10\\ 1 & 31 \end{pmatrix}$ \\ 
 & $=$ & 0.66 & 2.70 & $\begin{pmatrix} 8 & 6\\ 1 & 31 \end{pmatrix}$ \\ 
 & Minority oversampling & 0.74 & 2.06 & $\begin{pmatrix} 8 & 6\\ 1 & 31 \end{pmatrix}$ \\ 
\midrule
Random & $\propto$ & 0.71 & 3.66 & $\begin{pmatrix} 3 & 11\\ 0 & 32 \end{pmatrix}$ \\ 
 & $\frac{1}{\propto}$ & 0.69 & 3.65 & $\begin{pmatrix} 3 & 11\\ 0 & 32 \end{pmatrix}$ \\ 
 & $=$ & 0.71 & 2.64 & $\begin{pmatrix} 4 & 10\\ 0 & 32 \end{pmatrix}$ \\ 
 & Minority oversampling & 0.74 & 2.41 & $\begin{pmatrix} 5 & 9\\ 1 & 31 \end{pmatrix}$ \\ 
\midrule
 None & $\propto$ & 0.50 & 10.60 & $\begin{pmatrix} 0 & 14\\ 0 & 32 \end{pmatrix}$ \\ 
  & $\frac{1}{\propto}$ & 0.51 & 10.18 & $\begin{pmatrix} 0 & 14\\ 0 & 32 \end{pmatrix}$ \\ 
 & $=$ & 0.50 & 10.44 & $\begin{pmatrix} 0 & 14\\ 0 & 32 \end{pmatrix}$ \\ 
 & Minority oversampling & 0.47 & 11.09 & $\begin{pmatrix} 0 & 14\\ 0 & 32 \end{pmatrix}$ \\ 
\bottomrule\label{tab:negative_sampling}\end{tabularx}
\end{table}

\begin{table}
\caption{The effect of data augmentation on slide-level tasks using the K-means negative sampling strategy, 256 as the tile stride, and EfficientNet-B0 architecture.}
\begin{tabularx}{\linewidth}{X X X X X X X}
\toprule
Random crop & Stain norm. & Color jittering & Affine trs. & Dice coeff. & Error in $\sqrt{d_1 d_2}$ & Cfs. mtx. \\
\midrule
 &  &  &  & 0.72 & 2.58 & $\begin{pmatrix} 5 & 8 \\ 0 & 33 \end{pmatrix}$ \\ 
\cmark & \cmark &  &  & 0.73 & 2.03 & $\begin{pmatrix} 8 & 6 \\ 1 & 31 \end{pmatrix}$ \\ 
\cmark & & \cmark &  & 0.78 & 1.52 & $\begin{pmatrix} 7 & 7 \\ 1 & 31 \end{pmatrix}$ \\ 
\cmark & &  & \cmark & 0.78 & 2.22 & $\begin{pmatrix} 9 & 5\\ 1 & 31 \end{pmatrix}$ \\ 
\bottomrule
\label{tab:data_augmentation}\end{tabularx}
\end{table}

\begin{table}
\caption{The effect of architecture size/complexity on slide-level tasks. In all experiments, we use K-means negative sampling, 256 as the tile stride, and only random cropping as data augmentation. }
\begin{tabularx}{\linewidth}{X X X X}
\toprule
Architecture (\# params) & Dice coeff. & Error in $\sqrt{d_1 d_2}$ & Cfs. mtx. \\
\midrule
B0 (5M) & 0.74 & 2.06 & $\begin{pmatrix} 8 & 6\\ 1 & 31 \end{pmatrix}$ \\ 
B3 (12M) & 0.78 & 1.80 & $\begin{pmatrix} 9 & 5\\ 1 & 31 \end{pmatrix}$ \\ 
B5 (30M) & 0.71 & 1.74 & $\begin{pmatrix} 9 & 5\\ 2 & 30 \end{pmatrix}$ \\ 
B7 (58M) & 0.75 & 2.25 & $\begin{pmatrix} 9 & 5\\ 0 & 32 \end{pmatrix}$ \\ 
\bottomrule
\label{tab:model_complexity}\end{tabularx}
\end{table}

\begin{table}
\caption{The effect of sliding window tile stride on slide-level tasks using the EfficientNet-B3 architecture. In all experiments, we use K-means negative sampling and only random cropping as data augmentation. }
\begin{tabularx}{\linewidth}{X X X X}
\toprule
Stride & Dice coefficient & Error in $\sqrt{d_1 d_2}$ & Cfs. mtx. \\
\midrule
512 & 0.75 & 2.18 & $\begin{pmatrix} 2 & 12\\ 0 & 32 \end{pmatrix}$ \\ 
256 & 0.78 & 1.80 & $\begin{pmatrix} 9 & 5\\ 1 & 31 \end{pmatrix}$ \\ 
128 & 0.75 & 2.00 & $\begin{pmatrix} 11 & 3\\ 0 & 32 \end{pmatrix}$ \\ 
64 & 0.75 & 2.02 & $\begin{pmatrix} 10 & 4\\ 0 & 32 \end{pmatrix}$ \\ 
32 & 0.74 & 1.98 & $\begin{pmatrix} 9 & 5\\ 0 & 32 \end{pmatrix}$ \\ 
16 & 0.74 & 2.00 & $\begin{pmatrix} 9 & 5\\ 0 & 32 \end{pmatrix}$ \\ 
\bottomrule
\label{tab:stride_length}\end{tabularx}
\end{table}

\begin{table}
\caption{Comparison between references (expert pathologists), and our method on slide-level tasks using the EfficientNet-B3 architecture, including the consensus ground truth (with 8\% of the dataset removed). P\#* indicates the expert ID, and \textit{CNN} is the trained network to segment the tumor bed automatically. In all experiments, we use K-means negative sampling, 128 as the tile stride, and only random cropping as data augmentation.}
\begin{tabularx}{\linewidth}{X X X X X}
\toprule
Reference & Target & Dice coefficient & Error in $\sqrt{d_1 d_2}$ & Cfs. mtx. \\
\midrule
P\#1 & CNN & 0.74 & 2.10 & $\begin{pmatrix} 9 & 4\\ 1 & 32 \end{pmatrix}$ \\
& P\#2 & 0.78 & 2.11 & $\begin{pmatrix} 13 & 0\\ 0 & 33 \end{pmatrix}$ \\
& P\#3 & 0.80 & 1.63 & $\begin{pmatrix} 13 & 0\\ 0 & 33 \end{pmatrix}$ \\
\midrule
P\#2 & CNN & 0.73 & 2.58 & $\begin{pmatrix} 9 & 4\\ 1 & 32 \end{pmatrix}$ \\
 & P\#3 & 0.81 & 1.71 & $\begin{pmatrix} 13 & 0\\ 0 & 33 \end{pmatrix}$ \\
\midrule
P\#3 & CNN & 0.78 & 1.31 & $\begin{pmatrix} 9 & 4\\ 1 & 32 \end{pmatrix}$ \\
\midrule
Consensus & CNN & 0.75 & 2.00 & $\begin{pmatrix} 11 & 3\\ 0 & 32 \end{pmatrix}$ \\
& P\#1 & 0.81 & 1.46 & $\begin{pmatrix} 13 & 1\\ 0 & 32 \end{pmatrix}$ \\
& P\#2 & 0.84 & 1.35 & $\begin{pmatrix} 13 & 1\\ 0 & 32 \end{pmatrix}$ \\
& P\#3 & 0.82 & 1.45 & $\begin{pmatrix} 13 & 1\\ 0 & 32 \end{pmatrix}$ \\
\bottomrule
\label{tab:ref_target_consensus}\end{tabularx}
\end{table}

\begin{table}
\caption{Comparison between references (expert pathologists), and our method on slide-level tasks using the EfficientNet-B3 architecture, prior to including consensus ground truth. P\#* indicates the expert ID, and \textit{CNN} is the trained network to segment the tumor bed automatically. In all experiments, we use K-means negative sampling, 128 as the tile stride, and only random cropping as data augmentation.}
\begin{tabularx}{\linewidth}{X X X X X}
\toprule
Reference & Target & Dice coefficient & Error in $\sqrt{d_1 d_2}$ & Cfs. mtx. \\
\midrule
P\#1 & CNN & 0.71 & 2.49 & $\begin{pmatrix} 9 & 5\\ 2 & 34 \end{pmatrix}$ \\
& P\#2 & 0.74 & 2.86 & $\begin{pmatrix} 13 & 0\\ 1 & 36 \end{pmatrix}$ \\
& P\#3 & 0.78 & 2.00 & $\begin{pmatrix} 13 & 0\\ 1 & 36 \end{pmatrix}$ \\
\midrule
P\#2 & CNN & 0.71 & 2.89 & $\begin{pmatrix} 9 & 4\\ 2 & 35 \end{pmatrix}$ \\
 & P\#3 & 0.78 & 1.96 & $\begin{pmatrix} 13 & 0\\ 0 & 37 \end{pmatrix}$ \\
\midrule
P\#3 & CNN & 0.75 & 1.33 & $\begin{pmatrix} 9 & 4\\ 2 & 35 \end{pmatrix}$ \\
\bottomrule
\label{tab:ref_target_beforeconsensus}\end{tabularx}
\end{table}

\section{Discussion and conclusion}

8\% of the test set had to be discarded as the three pathologists were unable to reach consensus; this demonstrates the challenge associated with estimating the tumor bed, even for experts. In clinical practice, such challenging cases can be resolved by using an additional stain (e.g., immunostaining). Interestingly, two false positives and two false negative WSIs that our system initially detected were discarded in the consensus set, indicating errors made by the network are not random, and the failure points can be attributed to the uncertainty in the slide.

We find using randomly sampled negative instances to be particularly ineffective in identifying tumor-free slides. In effect, randomly sampling negative regions undersamples rare regions that exist in WSIs. In contrast, clustering negative samples based on a salient feature representation allows us to group visually similar instances. Therefore, clustering lets us sample the negative space more evenly, which improves precision. Furthermore, relying only on annotator labeled negative data (\textit{None} setting) is incapable of identifying true negatives and has a much lower performance compared to either K-means or random sampling. Upon examination of the predictions, we find that the network is incapable of identifying structures such as red blood cells or scanner imperfections (e.g., shadow effects), which are not present in the original BreastPathQ dataset, however negative sampling from Post-NAT BRCA WSIs provide us with this information. While it might not always be possible to obtain a boundary beyond which there is no cancer present, many datasets contain negative WSIs that can be sampled in a similar fashion.

After incorporating the negative samples, our dataset is highly imbalanced in favor of negative samples (85\%). A common strategy in handling class imbalance is to increase the cost of mislabeling the minority class by using weighted cross-entropy, where the weights per class are inversely proportional to the number of training samples with that label. We found that this strategy artificially skews the network to predict more samples from the minority class, which increases the number of false negatives. A counter-intuitive approach, where we use directly proportional weights, performs better, and not accounting for the imbalance (equal weights) gives the best performance out of these three approaches.  Finally, we observe that oversampling the minority class to rebalance class distributions gives the  best overall performance. Therefore, we use K-means negative samples with minority oversampling in slide-level experiments presented from Table \ref{tab:model_complexity} onward. In contrast, the results for the patch-level classification task shown in Table \ref{tab:patch_level_cls} suggest that any form of negative sampling adversely impacts prediction accuracy. Injecting information that is not present in the test set has a negative effect since the network has learned a more complex decision boundary that is not utilized for testing. Interestingly, random sampling is not as detrimental, possibly because most of the negative training instances look significantly different from positives.


We find that randomly cropping the instances improve results in both slide-level segmentation and classification tasks. We find that light color jittering performs better than stain normalization. Affine transformations do not improve results for the WSI task. In patch level analysis, we observe that augmentations are less significant, and that more than two test time augmentations (including random cropping by $448\times448$, flips and rotations) negatively impact the prediction accuracy. This is in contrast to WSI level predictions, where a smaller tile stride (corresponding to test-time augmentations in patches) is usually correlated with better performance. 

We find that EfficientNet-B3, which is comparable to ResNet34 and ResNet50 in terms of model complexity, provides the best bias and variance trade-off for our data and slide-level tasks. For the patch-level task, we find larger architectures to be more suitable, where the best performing network (B5) is 2.5$\times$ the size of best performing model (B3) for tumor bed estimation.

We find that stride lengths smaller than the patch length primarily help identify negative regions, and marginally improve the tumor bed outlines. We also find that stride lengths drastically smaller than the patch length do not lead to significantly benefit results, and are computationally prohibitive (e.g., a stride length of 16 processes each pixel 1024 times, and each WSI in this setting can have more than a million patches to be evaluated).

In conclusion, we find that best practices in patches versus WSI analysis vary significantly, despite employing the same networks with the same training data. This discrepancy also highlights that the advancements in either setting cannot be directly applied to the other, and independent research is required to further both fields.




\section*{Conflict of interest}

ALM is co-founder and CSO of Pathcore. Other authors have no conflict to declare.

\section*{Acknowledgments}

This research was enabled in part by support provided by Compute Canada (www.computecanada.ca). This work was funded by Canadian Cancer Society (grant \#705772) and NSERC RGPIN-2016-06283.

\appendix

\section{Estimating the dimensions from the tumor bed outline}\label{apx:tumor_bed_calc}

\begin{figure*}[!t]
     \centering
     \begin{subfigure}[b]{0.19\textwidth}
         \centering
         \includegraphics[width=\textwidth]{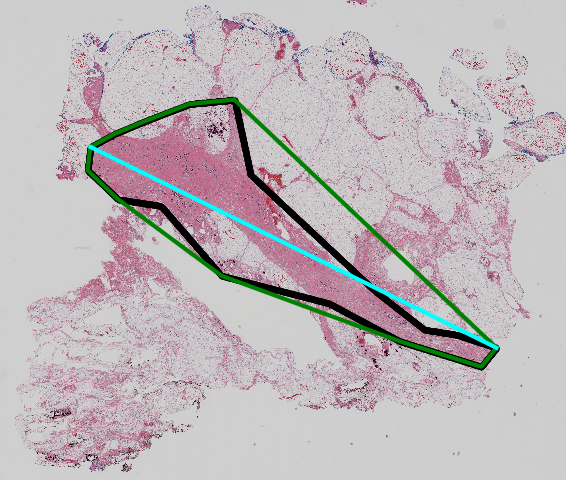}
         \caption{The longest diagonal (turquoise).}
         \label{fig:tumor_bed_axes_calc_a}
     \end{subfigure}
     \begin{subfigure}[b]{0.19\textwidth}
         \centering
         \includegraphics[width=\textwidth]{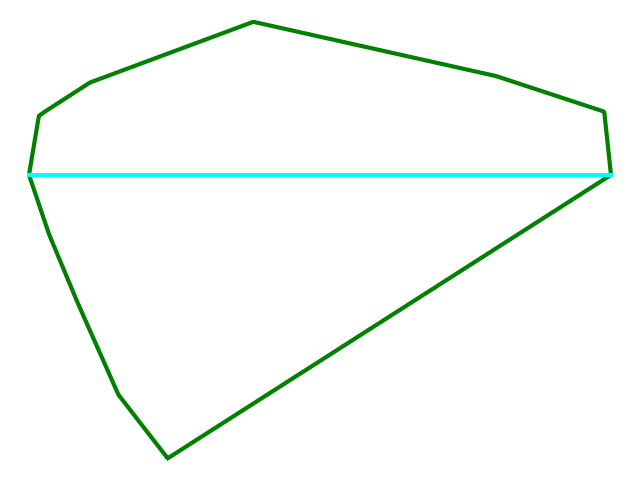}
         \caption{Rotating the hull for computational simplicity.}
         \label{fig:tumor_bed_axes_calc_b}
     \end{subfigure}
     \begin{subfigure}[b]{0.19\textwidth}
         \centering
         \includegraphics[width=\textwidth]{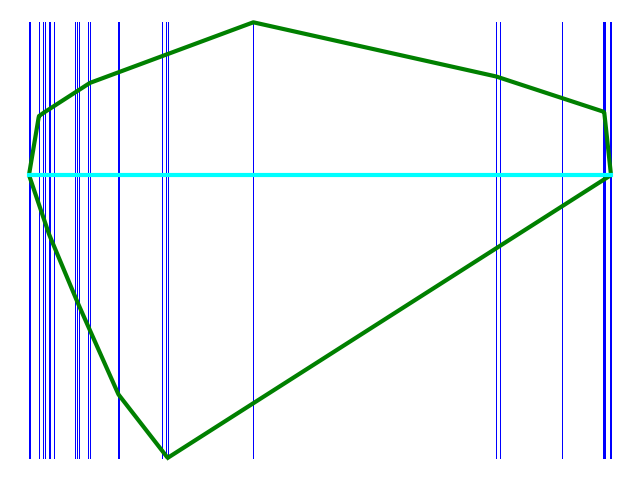}
         \caption{Drawing perpendicular lines to the diagonal.}
         \label{fig:tumor_bed_axes_calc_c}
     \end{subfigure}
     \begin{subfigure}[b]{0.19\textwidth}
         \centering
         \includegraphics[width=\textwidth]{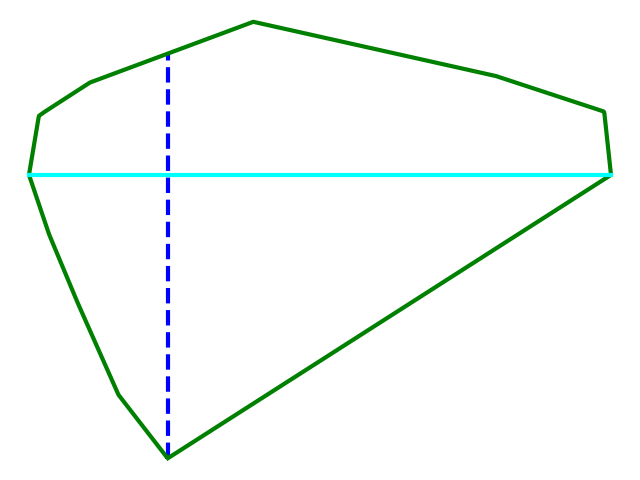}
         \caption{The longest line segment within the tumor bed.}\label{fig:tumor_bed_axes_calc_d}
     \end{subfigure}
     \begin{subfigure}[b]{0.19\textwidth}
         \centering
         \includegraphics[width=\textwidth]{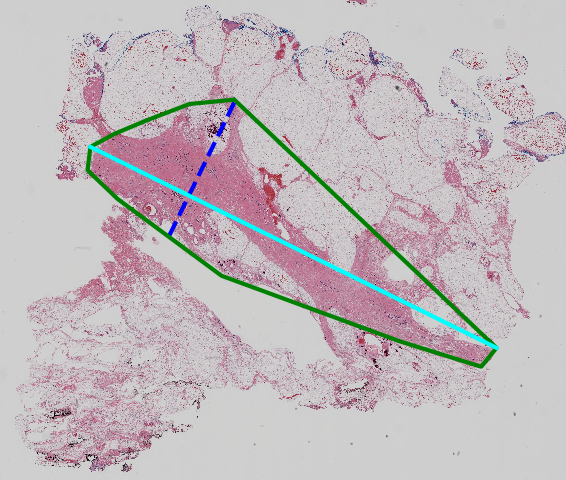}
         \caption{$d_1$ (turquoise) and $d_2$ (blue) for the tumor bed.}
         \label{fig:tumor_bed_axes_calc_e}
     \end{subfigure}
        \caption{Extraction of tumor bed extent, $d_1$ and $d_2$, from  the convex hull (shown in green on Fig. \ref{fig:tumor_bed_axes_calc_a}) of the tumor bed (shown in black on 
        Fig. \ref{fig:tumor_bed_axes_calc_a}).}\label{fig:tumor_bed_axes_calc}
\end{figure*}

As a quantitative measure of the tumor extent, \cite{symmans2007measurement} use the longest diagonal line $d_1$ inside the tumor bed, and the longest line $d_2$ within the tumor bed that is perpendicular to $d_1$, and combine them in a single quantity $d_{prim} = \sqrt{d_1 d_2}$. After calculating the tumor bed, we compute the convex hull as it simplifies the following calculations, since the convex shape has properties that are easier to deal with than of the tumor bed, which may be an irregular, concave polygon. Furthermore, we can apply $\mathcal{O}(n)$ time algorithms to estimate many geometric quantities (such as the diameter of a convex polygon) provided that we have the convex hull \citep{toussaint1983solving}. Given the convex hull, we first find the longest diagonal $d_1$ connecting two vertices of the convex hull (Fig. \ref{fig:tumor_bed_axes_calc_a}). Due to the properties of the convex hull, this diagonal is also the maximum extent in the tumor bed. Then, the convex hull is rotated along this longest diagonal using the 2d rotation matrix $R_{d_1}$ is defined as $\begin{bmatrix}
cos\theta & -sin\theta \\
sin\theta & cos\theta
\end{bmatrix}$ (Fig. \ref{fig:tumor_bed_axes_calc_b}). Given two endpoints of the longest diagonal $(x_1, y_1)$ and $(x_2, y_2)$, where the latter point is above the former ($y_2 > y_1$), $\theta$ is given by $-tan^{-1}(\frac{y_2 - y_1}{x_2-x_1})$. We use $\texttt{arctan2}$ function defined in most programming languages and packages for $tan^{-1}$. The resulting transformed hull is used to draw line segments perpendicular to the longest diagonal, crossing each vertex of the hull (Fig. \ref{fig:tumor_bed_axes_calc_c}). After the two crossing points (given that the vertex is not an extreme point on the convex hull) are identified, we pick the longest one as $d_2$ (Fig. \ref{fig:tumor_bed_axes_calc_d}), which is the longest line segment perpendicular to the original diagonal by the convex hull properties. The process is summarized in Fig. \ref{fig:tumor_bed_axes_calc}, where we overlaid the resulting dimensions on the original tumor bed for visualization.

\bibliographystyle{model2-names.bst}\biboptions{authoryear}
\bibliography{refs}

\end{document}